\documentclass[letter]{IEEEtran}

\usepackage{mathtools}
\usepackage{graphicx,color}
\usepackage{cite}
\usepackage{setspace} 
\usepackage{amsmath,amsthm,amssymb,bm}
\usepackage{multirow}
\usepackage{hhline}
\usepackage{epsfig}
\usepackage{epstopdf}
\usepackage{verbatim}
\usepackage{algorithm}
\usepackage{algpseudocode}
\usepackage{cases}
\usepackage{array}
\usepackage{subfigure}

\allowdisplaybreaks

\usepackage{forloop}
\newcounter{ct}

\makeatletter 
\def\@eqnnum{{\normalsize \normalcolor (\theequation)}} 
\makeatother
\begin{document}

\title{A Low-Complexity Post-Weighting Predistorter in a mMIMO Transmitter Under Crosstalk}
\author{Ganesh Prasad, \textit{Member, IEEE} and Håkan Johansson, \textit{Senior Member, IEEE}
	\thanks{G. Prasad and H. Johansson are with the Division of Communication Systems, Department of Electrical Engineering, Linköping University, 581 83 Linköping,
		Sweden (e-mail: \{ganesh.prasad, hakan.johansson\}@liu.se). 
}}

\maketitle
\begin{abstract}
	\label{R1C6} The beam-oriented digital predistortion (BO-DPD) is not sufficient to linearize the output from a subarray of power amplifiers (PAs) in different directions except the desired direction. Therefore, subsequent to the BO-DPD operation, we perform a post-weighting (PW) processing to minimize the nonlinear radiations in the wide range of directions under crosstalk. Here, the optimized PW coefficients are multiplied by the polynomial terms of the BO-DPD,	then, the resultant signals are distributed to the PAs to compensate the nonlinear radiations. In this work, first, we propose fully-featured post-weighting (FF-PW) scheme, then, we derive a low-complexity post-weighting (LC-PW) scheme.	
\end{abstract}   

\begin{IEEEkeywords}
	Predistortion, polynomial model, subarray of PAs,  post-weighting, convex optimization
\end{IEEEkeywords}
\IEEEpeerreviewmaketitle
\bstctlcite{IEEEexample:BSTcontrol}

\section{Introduction}\label{sec:intro}
For efficient transmission of signals, radio frequency (RF) power amplifiers (PAs) play an important role. However, the design of highly linear PAs over the large dynamic range of the signals is expensive. Further, it is costly for a massive multiple-input multiple-output (mMIMO) transmitter with a large number of PAs under crosstalk. So, the linearization requires a less complex and proficient predistortion scheme.

Initially, the multiple antennas transmitter was linearized using a single digital predistortion (DPD) by considering equal nonlinear characteristics of its PAs~\cite{yan19}. \label{page:R2C1} However, in practice, the nonlinear characteristics are not equal and later, instead of fully linearizing all PAs, the beam-oriented (BO) output in a desired direction was linearized using a single DPD, known as BO-DPD\footnote{In this work, the terms DPD and BO-DPD are used interchangeably.} under crosstalk~\cite{luo03,bri10}. But, it is not able to provide the linearization in other directions except the desired direction, thus, it gives nonlinear sidelobes in the BO output. Then, it is realized that a single DPD with one predistorted output signal is not sufficient to linearize all the PAs of different nonlinearites~\cite{ng06}. Later, the full linearization of the PAs was achieved by including a tuning box to each PA. But, it has high complexity as each tuning box requires a training~\cite{yu05}. 

Recently, in a post-weighting (PW) scheme, a single DPD training followed by a post-weighting (PW) optimization is used to generate more than one predistorted signal in a subarray, then, these signals are distributed to multiple PAs to address their nonlinearities~\cite{yan07}. But, it provides only one PW coefficient (one degree of freedom (DOF)) per PA that is not sufficient to linearize the multiple PAs. In the proposed PW scheme, the DOF per PA is increased with less complexity by reducing the adders, multipliers, and the RF chains. In this regard, the key contribution of this work is three-fold. (i) First, using the dual-input polynomial models of the PAs, an approximate relationship is established between the crosstalks at the PAs and the crosstalk compensation signal to the BO-DPD.  Then, using it, we train the DPD. (ii) Next, we propose a fully-featured PW (FF-PW) scheme and using it, we derive a low-complexity PW (LC-PW) scheme. Based on it, the system parameters are arranged non-trivially into suitable vectors and matrices to simplify the system analysis.  (iii)  Further, an expression for nonlinear radiation from the BO transmitter operating with DPD and PW is obtained. Using it, a convex minimization problem is formulated. Then, its optimal PW coefficients are obtained in a closed form. Finally, numerical results are obtained to get various design insights. 

\begin{figure}[!t] 
	\centering \includegraphics[width=3.2in]{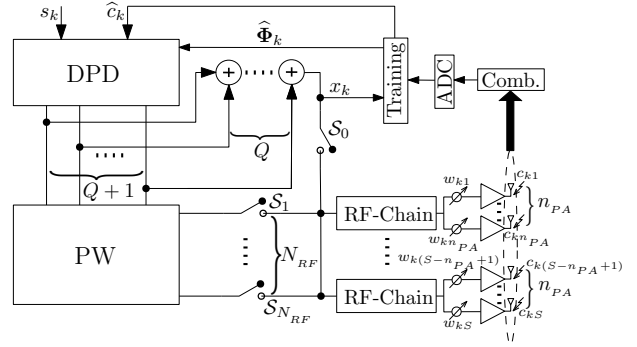}
	\caption{\small An architecture of a DPD followed by a PW predistortion.}    \label{fig:sys_mod}
\end{figure}

\section{System Description and BO-DPD Training}\label{sec:train}
\subsection{System Architecture}\label{sec:sys_arct}
\label{R1C11}In a mMIMO transmitter, we consider a uniform linear array of  $K \times S $ PAs where each of $K$ subarrays contains $S$ PAs. The message vector that needs to be transmitted is $\bm{s} =[s_1,\cdots, s_K]^T$, where $s_k$ is the message to be transmitted by the $k$th subarray. Fig.~\ref{fig:sys_mod} represents an architecture \label{page:R1C1IIa} for the linearization of the $k$th subarray under crosstalk signals $\{c_{kl}\}$; $l\in\{1,\cdots,S\}$ using two layers of predistortion operations: DPD followed by the PW predistortion. Here, $c_{kl}$ is the crosstalk signal at $l$th PA of the  $k$th subarray due to transmit signals from the remaining PAs. In this scheme, first, the DPD is trained separately for the BO output from the subarray. Thus, during the training of the DPD, the PAs are directly connected to the DPD and isolated from the PW block by closing the switch $\mathcal{S}_0$ and opening the switches, $\{\mathcal{S}_i\}$; $i\in\{1,\cdots,N_{_{RF}}\}$, where $N_{_{RF}}$ $(\le S)$ is the number of RF chains in the subarray. In the training, \label{page:R1C1IIb} the message $s_k$ and the signal $c_k$ are inputted to the dual-input DPD, where $c_k$ is obtained from the training block as shown in the figure. Using them, the DPD generates the predistorted signal $x_k$ to address the nonlinearties of the PAs under the crosstalks $\{c_{kl}\}$. Using Fig.~\ref{fig:sys_mod}, the generation of $x_k$ can be described as follows. The dual-input DPD is modeled using a polynomial model (as described later using~\eqref{eq:x_k}), having $Q+1$ terms which are its outputs. Each term is a basis function (of $s_k$ and $c_k$) multiplied by its coefficient. The $Q+1$ coefficients are represented by a vector $\bm{\Phi}_k$ which needs to be trained in the training block. The resultant predistorted signal from the DPD is $x_k$ which is the sum of $Q+1$ outputs (cf.~\eqref{eq:x_k} in Section~\ref{sec:sys_rep})\footnote{A conventional DPD \label{page:R1C1III} has one output $x_k$ based on a polynomial model~\cite{luo03}. But, here, each term of the polynomial is further processed by the PW block (cf. Fig.~\ref{fig:sys_mod}), thus, the DPD has $Q+1$ outputs and $x_k$ is the sum of them.}. Thereafter, $x_k$ is multiplied by the analog beamforming weights, $\{w_{kl}\}$ to get the BO output in a given direction, where $w_{kl}$ $(|w_{kl}|=1)$ is the analog phase shifter to the $l$th PA of the $k$th subarray and they are represented in a vector $\bm{w}_k=[w_{k1},\cdots,w_{kS}]^T$. The BO output is feedback to the training block to train the coefficients $\bm{\Phi}_k$ and the signal $c_k$ using the iterative method as described in Section~\ref{sec:train_dpd}. At the end of the training, their estimated values, $\bm{\widehat{\Phi}}_k$ and $\widehat{c}_k$ are obtained. Using them, the trained DPD provides the $Q+1$ outputs that are connected to the PAs via the PW block after opening $\mathcal{S}_0$ and closing $\{\mathcal{S}_i\}$. \label{page:R3C4a} In the PW block, the PW coefficients (dedicated for each $n_{_{PA}}=S/N_{_{RF}}$ PAs to enhance the predistortion DOF) are multiplied by the $Q$ nonlinear outputs of the DPD. Thereafter, the PW block outputs the $N_{_{RF}}$ number of signals and each of these signals are distributed to $n_{_{PA}}$ number of PAs to get the desired BO output.                                                                                                   

\subsection{Dual Input Polynomial Model of DPD and PAs}\label{sec:sys_rep}
We consider the dual-input memoryless polynomial models\footnote{For simplicity, we consider the memoryless polynomial models, however, the proposed work is equally applicable for memory polynomial models.}~\cite{luo03} for the the DPD and PAs. After omitting the nonlinear terms of the signal $c_k$, the output $x_k$ of the DPD as a function of $s_k$ and $c_k$ can be expressed as in~\eqref{eq:x_k} which is further represented in matrix form in~\eqref{eq:x_k_m}. 
\begin{subequations}
	\begin{align}\label{eq:x_k}
	\!\!\!x_k &=\hspace{-4mm} \sum_{p=0}^{(P-1)/2}\hspace{-4mm}\phi_{kp}^0\psi_p^0(s_k)+\hspace{-3mm}\sum_{p=0}^{(P-1)/2}\hspace{-4mm}\phi_{kp}^1\psi_p^1(s_k)c_k+\hspace{-3mm}\sum_{p=1}^{(P-1)/2}\hspace{-4mm}\phi_{kp}^2\psi_p^2(s_k)c_k^*,\!\!\!\!\\\label{eq:x_k_m}
	\!\!\!x_k &= \breve{\bm{\Psi}}(s_k,c_k)\breve{\bm{\Phi}}_k;\; k\in\{1,2,\cdots,K\},\!\!\!\!
	\end{align}
\end{subequations}
where $P$ is the order of the polynomial\footnote{Although, for simplicity, the orders of the polynomials are represented by the same symbol, $P$ in~\eqref{eq:x_k} and~\eqref{eq:y_kl}, they can take different values.} and $\breve{\bm{\Psi}}(s_k,c_k)=[\bm{\Psi}^0(s_k), \bm{\Psi}^1(s_k)c_k, \bm{\Psi}^2(s_k)c_k^*]$ is the row vector of basis functions, $\bm{\Psi}^v(s_k) =  [\psi_{\mu}^v(s_k), \cdots, $ $ \psi_{(P-1)/2}^v(s_k)]$ for $v\in\{0,1,2\}$.  $\psi_p^0(s_k)=s_k^{p+1}{s_k^*}^{p}$, $\psi_p^1(s_k)=s_k^p {s_k^*}^p$, and $\psi_p^2(s_k)=s_k^{p+1} {s_k^*}^{p-1}$. $\mu$ is the initial value of $p$ and $\mu=1$ for $v=2$; otherwise, $\mu=0$. $\breve{\bm{\Phi}}_k$ is a column vector of the coefficients for the basis functions in $\bm{\Psi}(s_k,c_k)$, given by $\breve{\bm{\Phi}}_k=[{\bm{\Phi}_k^0}^T, {\bm{\Phi}_k^1}^T, {\bm{\Phi}_k^2}^T]^T$. $\bm{\Phi}_k^v = [\phi_{k\mu}^v, \cdots, \phi^v_{k{(P-1)/2}}]^T$ and $\phi_{kp}^v$ is the coefficient of basis $\psi_p^v(s_k)$. Note that $\breve{\bm{\Psi}}(s_k,c_k)$ and $\breve{\bm{\Phi}}_k$ contain the basis functions and coefficients for the polynomial of order $P$ in~\eqref{eq:x_k}. However, in practice, certain basis functions with their nonzero coefficients play the dominant role in predistortion for a given type of PAs. Thus, in general, if $(Q+1)$ basis functions \label{page:R1C1IV} have their dominant role in the predistortion, then, the basis row vector $\bm{\Psi}(s_k,c_k)$ is defined as: $\bm{\Psi}(s_k,c_k) \triangleq [\psi_{p_1}^{v_1}\mathcal{C}_k^{v_1}, \cdots, \psi_{p_{(Q+1)}}^{v_{(Q+1)}}\mathcal{C}_k^{v_{(Q+1)}}]$ and corresponding coefficient column vector is: $\bm{\Phi}_k\triangleq[\phi_{kp_1}^{v_1}, \cdots, \phi_{kp_{(Q+1)}}^{v_{(Q+1)}}]^T$. Here, $\mathcal{C}_k^{v_i}=\delta(v_i)+c_k\delta(v_i-1)+c_k^*\delta(v_i-2)$, $p_i\in\{0,\cdots,(P-1)/2\}$, $v_i\in\{0,1,2\}$ for $i\in\{1,\cdots,(Q+1)\}$, $Q$ is the number of nonlinear basis functions, and only one basis function, $\psi_{p_i}^{v_i}\mathcal{C}_k^{v_i}=s_k$ is linear for $p_i=v_i=0$ (thus, the total number of basis functions is ($Q+1$)). $\delta(\cdot)$ is the Kronecker delta function. Hereafter, we consider the DPD in $\bm{\Psi}(s_k,c_k)$ and $\bm{\Phi}_k$.  Moreover, output signal vector of $K$ DPDs is denoted as $\bm{x}= [x_1,\cdots,x_K]^T$. 

Similarly,\label{page:R1C1V} based on dual input memoryless polynomial model, for the inputs, $x_k$ and $c_{kl}$  to the $l$th PA of the $k$th subarray, its output $y_{kl}$ is expressed as in~\eqref{eq:y_kl} and its matrix form in~\eqref{eq:y_kl_m}.
\begin{subequations}
	\begin{align}\nonumber\label{eq:y_kl}
	y_{kl} =\textstyle &\hspace{-0mm}\textstyle\sum_{p=0}^{(P-1)/2}\hspace{-1mm}\phi_{klp}^0\psi_p^0(w_{kl}x_k)+\hspace{-1mm}\sum_{p=0}^{(P-1)/2}\hspace{-0.5mm}\phi_{klp}^1\psi_p^1(w_{kl}x_{k})c_{kl}\\
	&\;+\hspace{-0mm}\textstyle\sum_{p=1}^{(P-1)/2}\hspace{-1mm}\phi_{klp}^2\psi_p^2(w_{kl}x_k)c_{kl}^*\\\label{eq:y_kl_m}
	y_{kl} = &\textstyle\breve{\bm{\Psi}}(w_{kl} x_k,c_{kl})\breve{\bm{\Phi}}_{kl};\; l\in\{1,2,\cdots,S\},		
	\end{align}
\end{subequations}
where $\breve{\bm{\Psi}}(w_{kl} x_k,c_{kl})$ and $\breve{\bm{\Phi}}_{kl}$ are the row and column vectors of the basis functions and its coefficients respectively. They can be represented similarly as described above for the DPD basis functions and coefficients with additional suffix $l$ to represent it for $l$th PA of the subarray. Further, for a given type of PAs, similar to the DPD, the $Q^{'}+1$ dominant basis functions to identify them can be represented in a row vector $\bm{\Psi}(w_{kl} x_k,c_{kl})$ and corresponding coefficients in a column vector can be given by $\bm{\Phi}_{kl}$. Note that the PAs coefficients $\bm{\Phi}_{kl}$; $l\in\{1,\cdots,S\}$ are assumed to be known which can be identified using least square (LS) estimation as described briefly in the numerical section. Further, all the outputs from the PAs can be expressed in a matrix form as:
\begin{align}
\bm{Y}_k = \bm{\Omega}_k\bm{\Theta}_k, \label{eq:Y_k}
\end{align}
where $\bm{Y}_k = [y_{k1},\cdots,y_{kS}]^T$, $\bm{\Omega}_k = \text{diag}([\bm{\Psi}(w_{k1} x_k,c_{k1})^T,$ $\cdots,$ $\bm{\Psi}(w_{kS} x_k,c_{kS})^T]^T)$ and $\bm{\Theta}_k = [\bm{\Phi}_{k1}^T,\cdots, \bm{\Phi}_{kS}^T]^T$. 

Using~\eqref{eq:Y_k}, for the stearing vector $\bm{h}_{k}^\varphi$ at angle (direction) $\varphi$ to the vertical plane of the array, the BO signal $z_{k}^\varphi$ is:
\begin{align}\label{eq:z_k}
z_{k}^\varphi = {\bm{h}_{k}^\varphi}^T\bm{Y}_k={\bm{h}_{k}^\varphi}^T\bm{\Omega}_k\bm{\Theta}_k,
\end{align}
where $\bm{h}_{k}^\varphi = [h_{k1}^\varphi,\cdots,h_{kS}^\varphi]^T$ and $h_{kl}^\varphi$ is the $l$th steering element. As $c_{kl}$ cannot be measured at the $l$th PA, it can be expressed as a linear combination of the transmit signals from the other PAs as~\cite{luo03}:
\begin{align}
c_{kl} =\textstyle\sum_{i=1}^{K}\sum_{r=1}^{S}\lambda_{kl,ir}^{'}y_{ir}= \bm{\lambda}_{kl}^{'}\bm{Y},\label{eq:c_kl}
\end{align}
where $\bm{\lambda}_{kl}^{'}=[\bm{\lambda}^{'}_{kl,1}, \cdots, \bm{\lambda}^{'}_{kl,K}]$, and $\bm{\lambda}^{'}_{kl,i}=[\lambda^{'}_{kl,i1}, \cdots, \lambda^{'}_{kl,iS}]$, and $\bm{Y}=[\bm{Y}_1^T,\cdots,\bm{Y}_K^T]^T$. $\lambda^{'}_{kl,ir}$ is the coefficient for contribution in crosstalk signal $c_{kl}$ at $l$th PA of the $k$th subarray from $r$th PA of the $i$th subarray. Simplifying~\eqref{eq:c_kl} in linear terms of $c_{ir}$ and $x_k$ after omitting the negligible nonlinear terms, we get $c_{kl}$ as in~\eqref{eq:c_kl_1}. Further, it is expressed in a matrix form in~\eqref{eq:c_matrix}, then, simplified to~\eqref{eq:c}.
\begin{subequations}
	\begin{align}\label{eq:c_kl_1}
	&c_{kl} =\textstyle  \sum_{i=1}^{K}\sum_{r=1}^{S}\big[\lambda_{kl,ir}^{'}\phi_{ir0}^0w_{ir}x_i+\lambda_{kl,ir}^{'}\phi_{ir0}^1c_{ir}\big]\\\label{eq:c_matrix}
	&\bm{c} = \textstyle \bm{A}^0W_D\bm{x} + \bm{A}^1\bm{c}\\\label{eq:c}&\Rightarrow
	\bm{c} = (I-\bm{A}^1)^{-1}\bm{A}^0W_D\bm{x} = \bm{\Lambda}W_D\bm{x},
	\end{align}
\end{subequations}
where $\bm{c}=[\bm{\overline{c}}_1^T,\cdots, \bm{\overline{c}}_K^T]^T$, $\bm{\overline{c}}_k=[c_{k1},\cdots,c_{kS}]^T$, $\bm{A}^v=\bm{\overline{\Lambda}}\text{diag}(\bm{\overline{\Phi}}_1^v,\cdots,\bm{\overline{\Phi}}_K^v)$, $\bm{\overline{\Phi}}_k^v=[\phi_{k10}^v,\cdots,\phi_{kS0}^v]$ for $v\in\{0,1\}$, $\bm{\overline{\Lambda}}=[\bm{\lambda}_{11}^{'T},\cdots,\bm{\lambda}_{1S}^{'T},\cdots,$ $\bm{\lambda}_{K1}^{'T},\cdots,\bm{\lambda}_{KS}^{'T}]^T$, and $W_D=\text{diag}([w_{11},\cdots,w_{1S}]^T,\cdots,[w_{K1},$ $\cdots,w_{KS}]^T)$. From~\eqref{eq:c}, $\bm{\Lambda}=(I-\bm{A}^1)^{-1}\bm{A}^0$ which is nothing but the coefficients associated with the weighted signal vector $W_D\bm{x}$ to get $\bm{c}$. It can be expressed as $\bm{\Lambda}= [\bm{\Lambda}_1^T,\cdots,\bm{\Lambda}_K^T]^T$, where $\bm{\Lambda}_k = [\bm{\lambda}_{k1}^T,\cdots,\bm{\lambda}_{kS}^T]^T$ and $\bm{\lambda}_{kl}=[\lambda_{kl,11},\cdots,\lambda_{kl,1S},\cdots, \lambda_{kl,K1},\cdots, \lambda_{kl,KS}]$.  Next, using \eqref{eq:z_k} and \eqref{eq:c}, we describe the training of the BO-DPD. 

\subsection{Training of BO-DPD}\label{sec:train_dpd}
In~\cite{luo03}, the relationship between $c_k$ and $\bm{\overline{c}}_k$ is obtained without considering $\bm{w}_k$. To include $\bm{w}_k$, first, we consider the same approximation, $\lambda_{kl,ir}\approx \alpha_{l}\lambda_{ki}$ due to uniform and linear arrangement of PAs~\cite{luo03}. It denotes that the crosstalk from $r$th PA of the $i$th subarray to $l$th PA of the $k$th subarray with coefficient $\lambda_{kl,ir}$ can be approximated to $\alpha_{l}$ times the overall crosstalk from the $i$th subarray to $k$th subarray with coefficient $\lambda_{ki}$. Applying it to $\bm{\lambda}_{kl}$ (cf. Section~\ref{sec:sys_rep}), we get: $\bm{\lambda}_{kl}\approx \alpha_l[\underbrace{\lambda_{k1},\cdots,\lambda_{k1}}_{S \text{ times}},\cdots,\underbrace{\lambda_{kK},\cdots,\lambda_{kK}}_{S \text{ times}}] = \alpha_{l}[\lambda_{k1}\bm{1}_S^T,\cdots,\lambda_{kK}\bm{1}_S^T]=\alpha_{l}\bm{\lambda}_k^TD_{\bm{1}}$, where $\bm{\lambda}_k= [\lambda_{k1},\cdots,\lambda_{kK}]^T$, $D_{\bm{1}}=\text{diag}(\underbrace{\bm{1}_S^T,\cdots,\bm{1}_S^T}_{K \text{ times}})$ and $\bm{1}_S$ is the column vector of ones of length $S$. Thus, $\bm{\Lambda}_k$ (cf. Section~\ref{sec:sys_rep}) is: $\bm{\Lambda}_k \approx \bm{\alpha} \bm{\lambda}_k^TD_{\bm{1}}$, where $\bm{\alpha}=[\alpha_{1},\cdots,\alpha_{S}]^T$. After applying this approximation in~\eqref{eq:c}, we get:
\begin{align}
\bm{\overline{c}}_k = \bm{\alpha}c_k;\;\;c_k=\bm{x}^TW_D^TD_{\bm{1}}^T\bm{\lambda}_k.\label{eq:c_k}
\end{align}
Using~\eqref{eq:c_k}, the inputs to the basis functions in $\bm{\Omega}_k$ of~\eqref{eq:Y_k} can be expressed in $x_k$ and $c_k$. As $c_k$ depends on $\bm{\lambda}_k$, we estimate $\bm{\lambda}_k$ to find $c_k$. In this regard, using~\eqref{eq:c_k}, $z_{k}^\varphi$ in~\eqref{eq:z_k} is given as:
\begin{align}
z_{k}^\varphi = g_k^0 + \bm{g}_k^1\bm{\widehat{\lambda}}_k + \bm{g}_k^2\bm{\widehat{\lambda}}_k^{*}, \label{eq:z_k_c_k}
\end{align}
where $g_k^0=\sum_{l=1}^{S}h_{kl}^\varphi\bm{\Psi}^0(w_{kl} x_k)\bm{\widehat{\Phi}}_{kl}^0$, $\bm{g}_k^1=\sum_{l=1}^{S}h_{kl}^\varphi$ $\bm{\Psi}^1(w_{kl} x_k)\bm{\widehat{\Phi}}_{kl}^1\bm{x}^TW_D^TD_{\bm{1}}^T$, and $\bm{g}_k^2 = \sum_{l=1}^{S}h_{kl}^\varphi\bm{\Psi}^2(w_{kl} x_k)\bm{\widehat{\Phi}}_{kl}^2\bm{x}^HW_D^HD_{\bm{1}}^H$. Further, by including time samples of $\bm{x}$, \eqref{eq:z_k_c_k} is given by:
\begin{align}
\bm{z}_{k}^\varphi = \bm{g}_k^0 + \bm{G}_k^1\bm{\widehat{\lambda}}_k + \bm{G}_k^2\bm{\widehat{\lambda}}_k^{*},\label{eq:z_k_ext_c_k}
\end{align}
where $z_{k}^\varphi$, $g_k^0$, $\bm{g}_k^1$, and $\bm{g}_k^2$ are denoted as $\bm{z}_{k}^\varphi$, $\bm{g}_k^0$, $\bm{G}_k^1$, and $\bm{G}_k^2$ respectively after including the time samples\footnote{Note that in this work, the boldface symbol $\bm{a}$ represents the scalar signal $a$ with its time samples.}. By splitting~\eqref{eq:z_k_ext_c_k} into real and imaginary parts, the real part ${\mathcal{R}}(\bm{\widehat{\lambda}}_k)$ and the imaginary part $\mathcal{I}(\bm{\widehat{\lambda}}_k)$, can be determined as:
\begin{eqnarray}\label{eq:reim_lam}
\left[ \hspace{-2mm}\begin{array}{c}
\mathcal{R}(\bm{\widehat{\lambda}}_k)\\ \mathcal{I}(\bm{\widehat{\lambda}}_k)
\end{array} \hspace{-2mm}\right]\hspace{-1mm} =\hspace{-1mm}
\left[\hspace{-2mm}\begin{array}{cc}
\mathcal{R}(\bm{G}_k^1 \hspace{-1mm}+ \hspace{-1mm}\bm{G}_k^2) & \mathcal{I}(-\bm{G}_k^1 \hspace{-1mm}+\hspace{-1mm} \bm{G}_k^2) \\ \mathcal{I}(\bm{G}_k^1 \hspace{-1mm}+\hspace{-1mm} \bm{G}_k^2) & \mathcal{R}(\bm{G}_k^1\hspace{-1mm} - \hspace{-1mm}\bm{G}_k^2)
\end{array}\hspace{-2mm}\right]^{\dagger}
\hspace{0mm} \left[\hspace{-2mm}\begin{array}{c}
\mathcal{R}(\bm{z}_{k}^\varphi\hspace{-1mm}-\hspace{-1mm}\bm{g}_k^0) \\ \mathcal{I}(\bm{z}_{k}^\varphi\hspace{-1mm}-\hspace{-1mm}\bm{g}_k^0)
\end{array}\hspace{-2mm}\right]
\end{eqnarray}

\begin{figure}[!t] 
	\centering  \includegraphics[width=2.6in]{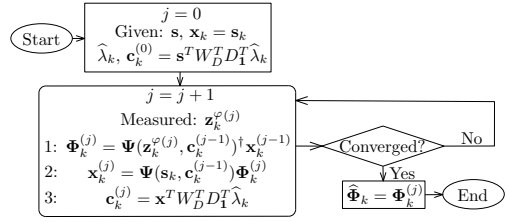}
	\caption{\small Flow diagrams for the identification of DPD coefficients.}    \label{fig:flow_charts}
\end{figure}
\noindent To determine $\bm{\widehat{\lambda}}_k$, first, $\bm{x}_k$ is set as: $\bm{x}_k=\bm{s}_k$. Thus, we can compute $\bm{G}_k^1$, $\bm{G}_k^2$, and $\bm{g}_k^0$ and using the measured $\bm{z}_{k}^\varphi$, from \eqref{eq:reim_lam}, we find $\bm{\widehat{\lambda}}_k$. Further, the obtained $\bm{\widehat{\lambda}}_k$ is used in the algorithm as shown in Fig.~\ref{fig:flow_charts}. Here, initially, we again set $\bm{x}_k = \bm{s}_k$ and for given $\bm{\widehat{\lambda}}_k$, $c_k$ is determined. Then, using post-inverse, the measured $\bm{z}_k^\varphi$ and $c_k$ are set as inputs to the DPD  to find $\bm{\Phi}_k$ using LS method. Using it, the output $\bm{x}_k$ of the DPD and $\bm{c}_k$ are computed. The process repeats until the value of  $\bm{\Phi}_k$ converges. The complexity \label{page:R2C4} of the algorithm in an iteration is determined using the dominant matrix operations in Steps~1, 2, and 3 (cf. Fig.~\ref{fig:flow_charts}) which is: $O((Q+1)^2)+O(K^2S)$.

\begin{figure}[!t]
	\centering
	\subfigure[]{\includegraphics[width=3.0in]{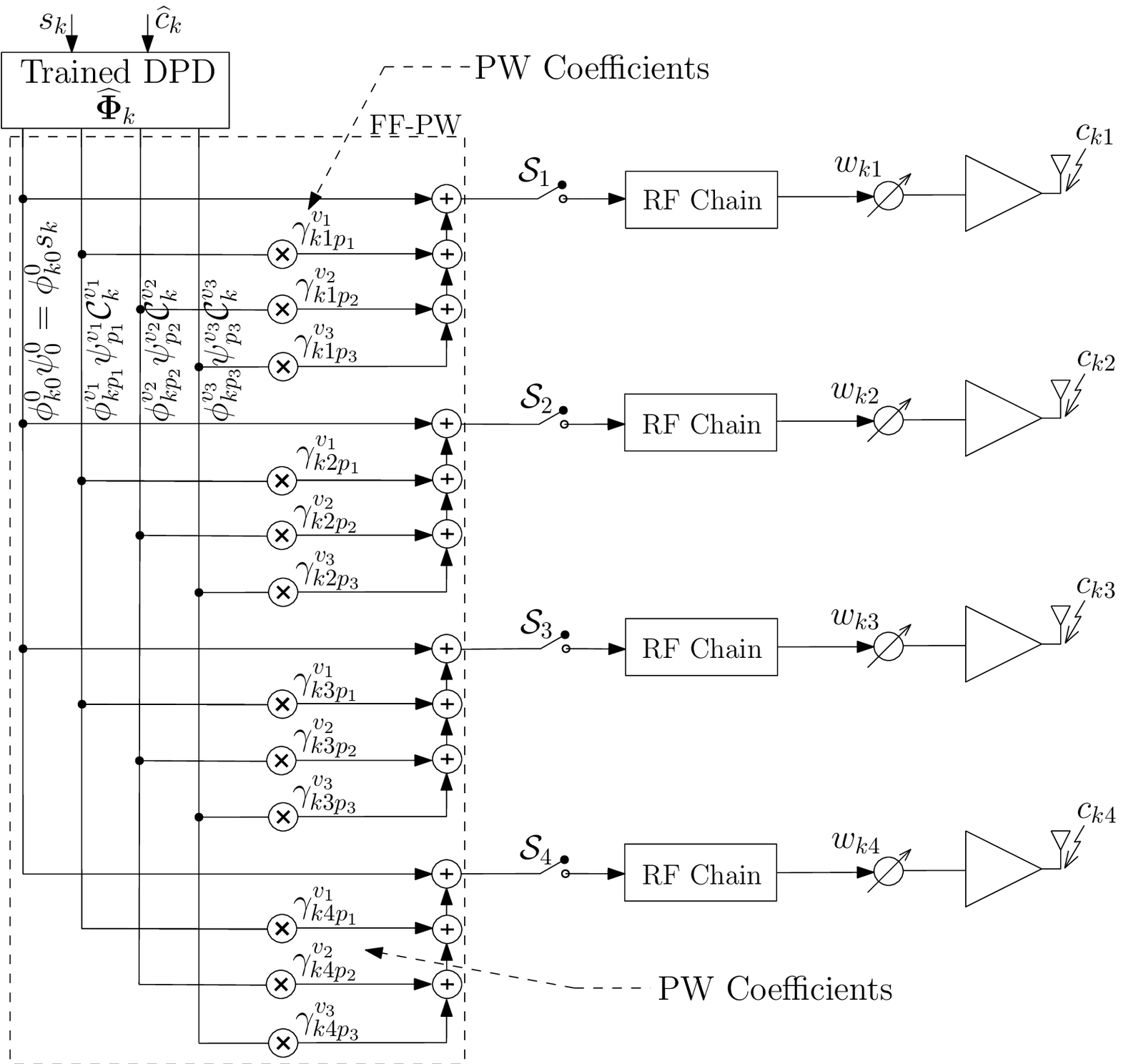}}
	\subfigure[]{\includegraphics[width=3.0in]{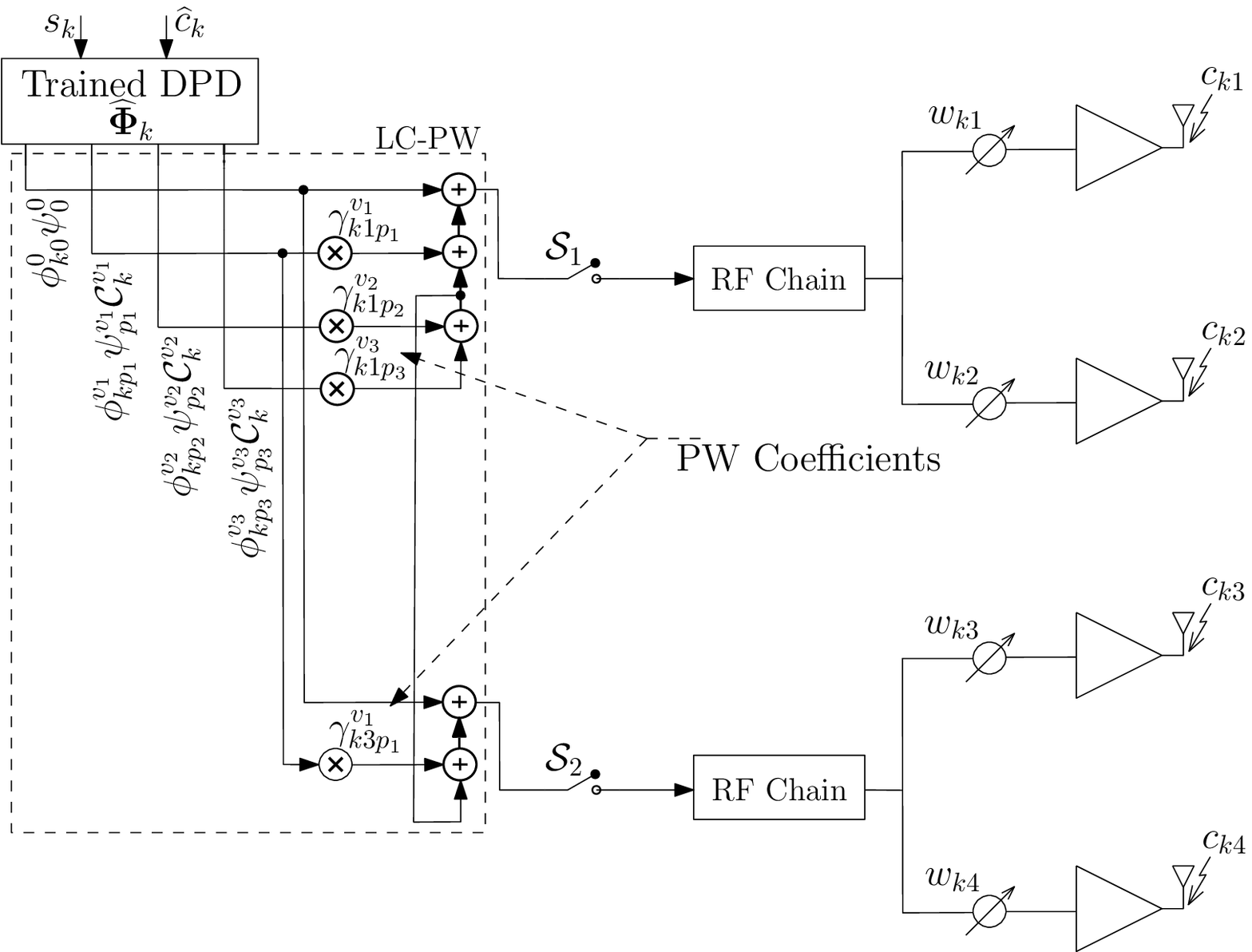}}
	\caption{\small Two PW schemes, (a) FF-PW and (b) LC-PW for $S=4$.}
	\label{fig:PW_Arch}
\end{figure}
\section{Post-Weighting Schemes and Optimization}\label{sec:PW}
To investigate \label{page:R2C2} the PW processing, first, we express $z_k^\varphi$ in~\eqref{eq:z_k_c_k} as a function of $s_k$. By substituting~\eqref{eq:x_k_m} into~\eqref{eq:z_k_c_k} and separating the first basis function from the rest, we get: 
\begin{align}\label{eq:z_k_NL}
\!\!\!z_k^\varphi \hspace{-1.1mm} = \hspace{-0.7mm} {\bm{h}_k^\varphi}^T\hspace{-0.4mm}\bm{W}_k \bm{\tilde{\phi}}_{k}^0\phi_{k0}^0s_k \hspace{-1mm}+ \hspace{-0.9mm}{\bm{h}_{k0}^\varphi}^T\hspace{-0.6mm}\bm{W}_k \bm{\tilde{\phi}}_{k0}^0\hspace{-0.5mm}\bm{\Psi}^{'}\hspace{-0.8mm}(\hspace{-0.7mm}s_k,c_k\hspace{-0.7mm})\bm{\Phi}^{'}_k\hspace{-1mm} +\hspace{-1mm} \tilde{Z}_{k,NL}^\varphi,\!\!\!\!\!
\end{align}
where $\bm{W}_k=\text{diag}({\bm{w}_k})$, $\bm{\tilde{\phi}}_{k0}^0 = [\phi_{k10}^0, \cdots, \phi_{kS0}^0]^T$, $\bm{\Psi}^{'}(s_k,c_k)= \{\bm{\Psi}(s_k,c_k)\backslash \psi_0^0(s_k)=s_k\}$,  $\bm{\Phi}^{'}_k= \{\bm{\Phi}_k\backslash \phi_{k0}^0\}$, and $\tilde{Z}_{k,NL}$ is the nonlinear higher order terms with less power contents, obtained after removing the term consisting first (linear) basis function in~\eqref{eq:z_k}. In~\eqref{eq:z_k_NL}, the first term is the desired output. But, the second and third terms inject nonlinearties, denoted as the nonlinear radiation $z_{k,NL}^\varphi$ as:
\begin{align}\label{eq:z_NL}
z_{k,NL}^\varphi = {\bm{h}_{k}^\varphi}^T\bm{W}_k \bm{\tilde{\phi}}_{k0}^0\bm{\Psi}^{'}(s_k,c_k)\bm{\Phi}^{'}_k + \tilde{Z}_{k,NL}^\varphi.
\end{align}
\noindent In~\eqref{eq:z_NL}, $z_{k,NL}^\varphi$ gives the nonlinearties in other directions except $\varphi$. Because, in BO-DPD, its coefficients $\bm{\Phi}^{'}_k$ are trained to provide the linearization in the direction $\varphi$. Therefore, to linearize it further in other directions too, the predistortion output of the trained DPD is passed through the PW block by changing the modes of the switches ($\mathcal{S}_0$ is open and $\{\mathcal{S}_i\}$ are closed). Next, we describe the proposed two PW schemes. 

\subsection{Post-Weighting}\label{sec:post_weight}
The outputs of the DPD \label{page:R1C1Ib1} are the $Q+1$ basis functions (in $\bm{\Psi}(s_k,c_k)$) multiplied by the respective coefficients (in $\bm{\Phi}_k$) and they are inputted to the PW block. The two types of PW schemes: (a) FF-PW and (b) LC-PW are shown in Fig.~\ref{fig:PW_Arch} for $S=4$ and $Q=3$. The PW coefficients are only multiplied by the $Q$ nonlinear outputs of the DPD comprising the nonlinear basis functions (in $\bm{\Psi}^{'}(s_k,c_k)$) except the linear basis function $\psi_{p_i}^{v_i}\mathcal{C}_k^{v_i}=s_k$ for $p_i=v_i=0$ (cf. Section~\ref{sec:sys_rep}). Thereafter, the outputs of the PW block are distributed to the inputs of the PAs to get the desired linear output. 

\subsubsection{Fully-Featured PW (FF-PW)}\label{sec:fully-featured} 
For instance, the \label{page:R1C1Ib2} FF-PW scheme in Fig.~\ref{fig:PW_Arch}(a) is for $S=4$ and $Q=3$. Here, $Q=3$ nonlinear outputs of the DPD are multiplied by the $S=4$ different sets of PW coefficients to generate the resultant predistorted signals for the respective $S$ PAs. For the generation of the predistorted signal to the $l$th $(l\in \{1,2,3,4\})$ PA, the corresponding set of coefficients is $\{\gamma_{klp_1}^{v_1},\gamma_{klp_2}^{v_2},\gamma_{klp_3}^{v_3}\}$. These PW coefficients are multiplied by the respective DPD outputs given in the set $\{\phi_{kp_1}^{v_1}\psi_{p_1}^{v_1}\mathcal{C}_k^{v_1},\phi_{kp_2}^{v_2}\psi_{p_2}^{v_2}\mathcal{C}_k^{v_2},\phi_{kp_3}^{v_3}\psi_{p_3}^{v_3}\mathcal{C}_k^{v_3}\}$. Thereafter, the sum of the three multiplications along with the linear DPD output, $\phi_{k0}^0s_k$ gives the $l$th predistorted signal. Therefore, the total number of PW coefficients, $N_{\gamma}^{F}=S\times Q=12$ which are completely different from each other without any repetition and it provides the full DOF to further linearize the individual PA. However, in LC-PW, the PW coefficients are assigned adaptively (some coefficients repeat) to the PAs to reduce their number as described later. In general, the $Q$ PW coefficients associated with the predistorted signal to the $l$th PA are represented as: $\gamma_{klp_1}^{v_1},\cdots,\gamma_{klp_Q}^{v_Q}$, where $\gamma_{klp_i}^{v_i}$ is the PW coefficient multiplied by the $i$th nonlinear DPD output. Further, for the convenience in analysis, the PW coefficients of the $S$ PAs are arranged in a vector as: $\bm{\gamma}_k\triangleq [\gamma_{k1p_1}^{v_1},\cdots,\gamma_{kSp_1}^{v_1},\cdots, \gamma_{k1p_Q}^{v_Q},\cdots,\gamma_{kSp_Q}^{v_Q}]^T$, where the coefficients multiplied by the same DPD output are grouped together. The PW processing can be analyzed by incorporating the $\bm{\gamma}_k$ in~\eqref{eq:z_NL}. As its first term has $\bm{\Psi}^{'}(s_k,c_k)$, to multiply $\bm{\gamma}_k$ with it, we need to rearrange $\bm{\Psi}^{'}(s_k,c_k)$ along with other matrices and vectors according to arrangement of PW coefficients in $\bm{\gamma}_k$. So, $\bm{h}_{k}^\varphi$, $\bm{W}_k$, $\bm{\tilde{\phi}}_{k0}^0$, $\bm{\Psi}^{'}(s_k,c_k)$, and $\bm{\Phi}^{'}_k$ are arranged as $\bm{\overline{h}}_{k}^\varphi$, $\bm{\overline{W}}_k$, $\bm{\overline{\tilde{\phi}}}_{k0}^0$, $\bm{\overline{\Psi^{'}}}(s_k,c_k)$, and $\bm{\overline{\Phi^{'}}}_k$, respectively. $\bm{\overline{h}}_{k}^\varphi = \bm{1}_Q\otimes \bm{h}_k^\varphi$, $\bm{\overline{W}}_k=\text{diag}(\bm{1}_Q\otimes \bm{w}_k)$, $\bm{\overline{\tilde{\phi}}}_{k0}^0 = \text{diag}(\bm{1}_Q\otimes \bm{\tilde{\phi}}_{k0}^0)$, $\bm{\overline{\Psi^{'}}}(s_k,c_k) = \text{diag}(\bm{\Psi}^{'}(s_k,c_k)^T \otimes \bm{1}_S)$, and $\bm{\overline{\Phi^{'}}}_k = \text{diag}(\bm{\Phi}^{'}_k \otimes \bm{1}_S)$.  Now,~\eqref{eq:z_NL} can be expressed as:
\begin{align}\label{eq:z_NL_mod}
\overline{z}_{k,NL}^\varphi = T_k^{\varphi}\bm{\gamma}_k + Z_{k,NL}^\varphi,
\end{align}
where $T_k^{\varphi} = {\bm{\overline{h}}_{k}^\varphi}^T\bm{\overline{W}}_k \bm{\overline{\tilde{\phi}}}_{k0}^0\bm{\overline{\Psi^{'}}}(s_k,c_k)\bm{\overline{\Phi^{'}}}_k$. Also, after incorporating PW coefficients in $\tilde{Z}_{k,NL}^\varphi$, it is expressed as $Z_{k,NL}^\varphi$ in~\eqref{eq:z_NL_mod}. As described earlier, due to small power content, we neglect the PW effect on  $Z_{k,NL}^\varphi$. Moreover, in FF-PW, the number of multipliers $N_{\gamma}^F$ and the number of adders $N_a^F$ are same as the number of PW coefficients in $\bm{\gamma}_k$, given as: $N_{\gamma}^F=N_a^F=S\times Q$. Also, the number of RF chains $N_{RF}^F=S$. So, to reduce them, we propose a LC-PW scheme.

\subsubsection{Low-Complexity PW (LC-PW)}\label{sec:hybrid}
As the complexity of the PW block depends on the number of PW coefficients, \label{page:R1C1Ib3} therefore, in the LC-PW scheme, the number of PW coefficients are reduced by a factor based on a geometric sequence. In general, the nonlinear DPD outputs in this scheme are arranged in their decreasing order of dominance (or increasing order of the polynomial terms). Besides, the numbers of PW coefficients that are multiplied by the $Q$ DPD outputs, decrease in a geometric sequence as: $\{Sr^\nu, Sr^{(\nu+1)},\cdots, Sr^{(\nu+Q-1)}\}$, where $r$ $(<1)$ is the common ratio.  For example, in Fig.~\ref{fig:PW_Arch}(b), for $S=4$, $Q=3$, $r=1/2$, and $\nu=1$, the sequence of numbers of PW coefficients is: $\{2,1,0.5\}$. But, the number cannot be a fraction value, so, the value in the sequence less than one is assigned as one, thus the sequence is: $\{2,1,1\}$. So, we keep decreasing the number of coefficients until $Sr^i < 1$; $i\in\{\nu,\nu+1,\cdots,\nu+Q-1\}$ and after that we assign one PW coefficient to each of the remaining DPD outputs. Thus, in general, the total number of coefficients (or the multipliers),  $N_\gamma^L$ in the  LC-PW is given by~\eqref{eq:N_gamma} and the total number of adders, $N_a^L$ is expressed in~\eqref{eq:N_adder}.
\begin{subequations}
	\begin{eqnarray}\label{eq:N_gamma}
	&\hspace{-6mm}N_\gamma^L\hspace{-1mm} =\hspace{-1mm}
	\left\{\hspace{-2.5mm}\begin{array}{cc}
	S\times \frac{r^\nu(1-r^Q)}{1-r}; \!\!\!\!\!& \text{for } S\times r^{(Q+\nu-1)}\ge 1 \\ S\times \frac{r^\nu(1-r^m)}{1-r}\hspace{-0.5mm}+\hspace{-0.5mm}Q\hspace{-0.5mm}-\hspace{-0.5mm}m;\!\!\!\!\!& \text{for } \{S\times r^{(m+\nu-1)}\ge 1\}\\\!\!\!\!\!&\wedge\{S\times r^{(m+\nu)}< 1\},
	\end{array}\hspace{-2mm}\right.\\\label{eq:N_adder}
	&\hspace{-1.55in}N_a^L=N_\gamma^L+Sr^\nu-\left\lceil Sr^{(\nu+Q-1)} \right\rceil.
	\end{eqnarray}
\end{subequations}
For the first case in~\eqref{eq:N_gamma}, the PW coefficient vector, $\bm{\gamma}_k = [\gamma_{k1p_1}^{v_1},\gamma_{k(1+r^{-\nu})p_1}^{v_1},\cdots,\gamma_{k(1+(Sr^{\nu}-1)r^{-\nu}) p_1}^{v_1},\cdots, \gamma_{k1p_Q}^{v_Q},$ $\gamma_{k(1+r^{-(\nu+Q-1)})p_Q}^{v_Q},\cdots,\gamma_{k(1+(Sr^{(\nu+Q-1)}-1)r^{-(\nu+Q-1)})p_Q}^{v_Q}]^T$ and  $\bm{\overline{\Phi^{'}}}_k = \text{diag}(\bm{\Phi}^{'}_k \otimes \bm{1}_S)\text{diag}(\underbrace{\bm{1}_{r^{-\nu}},\cdots, \bm{1}_{r^{-\nu}}}_{Sr^\nu \text{ times}}, \cdots,$ $ \underbrace{\bm{1}_{r^{-(\nu+Q-1)}},\cdots, \bm{1}_{r^{-(\nu+Q-1)}}}_{Sr^{(\nu+Q-1)} \text{ times}})$. Otherwise, for the second case, $\bm{\gamma}_k = [\gamma_{k1p_1}^{v_1}, $ $  \gamma_{k(1+r^{-\nu})p_1}^{v_1},\cdots,\gamma_{k(1+(Sr^{\nu}-1)r^{-\nu}) p_1}^{v_1},\cdots, \gamma_{k1p_m}^{v_m},$ $\gamma_{k(1+r^{-(\nu+m-1)})p_m}^{v_m},\cdots,\gamma_{k(1+(Sr^{(\nu+m-1)}-1)r^{-(\nu+m-1)})p_m}^{v_m},$ $ \gamma_{k1p_{(m+1)}}^{v_{(m+1)}},$ $\gamma_{k1p_{(m+2)}}^{v_{(m+2)}},\cdots,\gamma_{k1p_Q}^{v_Q}]^T$ and $\bm{\overline{\Phi^{'}}}_k = \text{diag}(\bm{\Phi}^{'}_k \otimes \bm{1}_S) \text{diag}(\underbrace{\bm{1}_{r^{-\nu}},\cdots, \bm{1}_{r^{-\nu}}}_{Sr^\nu \text{ times}}, $ $ \cdots, \underbrace{\bm{1}_{r^{-(\nu+m-1)}},\cdots, \bm{1}_{r^{-(\nu+m-1)}}}_{Sr^{(\nu+m-1)} \text{ times}}, $ $ \underbrace{\bm{1}_S, \cdots, \bm{1}_S}_{Q-m \text{ times}})$. Moreover, remaining vectors and matrices, $\bm{\overline{h}}_{k}^\varphi$, $\bm{\overline{W}}_k$, and $\bm{\overline{\tilde{\phi}}}_{k0}^0$, $\bm{\overline{\Psi^{'}}}(s_k,c_k)$ are expressed same as for FF-PW (cf. Section~\ref{sec:fully-featured}). Also, the expression for the nonlinear radiation, $\overline{z}_{k,NL}^\varphi$ is same as in~\eqref{eq:z_NL_mod}. Moreover, from Fig.~\ref{fig:PW_Arch}(b), the number of RF chains, $N_{RF}^L$ depends on the number of coefficients assigned to the first nonlinear output, i.e., $N_{RF}^L=Sr^\nu=N_{RF}^Fr^\nu$. Thus, in LC-PW, the number of RF chains, multipliers, and the adders are reduced by the factors, $r^{-\nu}$, $N_\gamma^F/N_{\gamma}^L$, and $N_a^F/N_a^L$, respectively. For example, in Fig.~\ref{fig:PW_Arch}(b), the respective factors are $2$, $4$, and $2.4$. Hence, the LC-PW is less complex and economical for the mMIMO.

\subsection{Optimization of $\bm{\gamma}_k$}\label{sec:opt_gamma}
The optimization problem can be formulated as:
\begin{align}\nonumber
&\mathcal{P}_0\!:\; \underset{\bm{\gamma}_k}{\text{minimize}}\;\;\textstyle \sum_{t}
\hspace{-0mm}  \textstyle{\mathbb{E}[|\overline{z}_{k,NL}^{\varphi_t}|^2]} \\\nonumber
& \text{s. t.:}\;
\textstyle {\bm{\overline{h}}_{k}^{\varphi_0}}^T\bm{\overline{W}}_k \bm{\overline{\tilde{\phi}}}_{k0}^0\bm{\overline{\Psi^{'}}}\;\bm{\overline{\Phi^{'}}}_k\bm{\gamma}_k = {\bm{\overline{h}}_{k}^{\varphi_0}}^T\bm{\overline{W}}_k \bm{\overline{\tilde{\phi}}}_{k0}^0\bm{\overline{\Psi^{'}}}\;\bm{\overline{\Phi^{''}}}_k,
\end{align}
where $\bm{\overline{\Phi^{''}}}_k =\bm{\Phi}^{'}_k \otimes \bm{1}_S$ and $\mathbb{E}[\cdot]$ is the expectation with respect to time samples. In problem $\mathcal{P}_0$, the objective function which needs to be minimized in $\bm{\gamma}_k$, is the sum of the average value of the power of nonlinear radiation in the given range of directions with sample points $\{\varphi_t\}$. The constraint ensures the linearization of BO signal in the desired direction $\varphi_0$.

\subsubsection{Optimal Solution}\label{sec:opt_sol}
To investigate the convexity of the problem, the objective function, $\mathcal{O}_{\mathcal{P}_0}\triangleq\sum_{t}\mathbb{E}[|\overline{z}_{k,NL}^{\varphi_t}|^2] = \sum_{t}\mathbb{E}[{\overline{z}_{k,NL}^{\varphi_tH}}\;\overline{z}_{k,NL}^{\varphi_t}]$, can be expressed using~\eqref{eq:z_NL_mod} as:
\begin{align}\nonumber\label{eq:obj_fun}
\mathcal{O}_{\mathcal{P}_0} =& \textstyle \bm{\gamma}_k^H\sum_{t}\mathbb{E}[{T_k^{\varphi_t}}^HT_k^{\varphi_t}]\bm{\gamma}_k + \bm{\gamma}_k^H\sum_{t}\mathbb{E}[{T_k^{\varphi_t}}^HZ_{k,NL}^{\varphi_t}]\\
& + \textstyle \sum_{t}\mathbb{E}[{Z_{k,NL}^{\varphi_tH}}\;T_k^{\varphi_t}]\bm{\gamma}_k + \sum_{t}\mathbb{E}[{Z_{k,NL}^{\varphi_tH}}Z_{k,NL}^{\varphi_t}].
\end{align}
As the constraint is linear and from~\eqref{eq:obj_fun}, the objective function is quadratic in $\bm{\gamma}_k$, the problem $\mathcal{P}_0$ is convex and gives a global solution using the Karush–Kuhn–Tucker (KKT) conditions. From $\mathcal{P}_0$, the Lagrangian function, $\mathcal{L}(\bm{\gamma}_k,\eta)$ is: 
\vspace{-1mm}
\begin{align}
\!\!\!\!\mathcal{L}(\bm{\gamma}_k,\eta) = \mathcal{O}_{\mathcal{P}_0} + \eta(T_k^{\varphi_0}\bm{\gamma}_k-{\bm{\overline{h}}_{k}^{\varphi_0}}^T\bm{\overline{W}}_k \bm{\overline{\tilde{\phi}}}_{k0}^0\bm{\overline{\Psi^{'}}}\;\bm{\overline{\Phi^{''}}}_k),
\end{align}
where $\eta$ is the Lagrangian multiplier which is $\ne 0$ to consider the constraint in the optimization. Using the complex gradient of $\mathcal{L}(\bm{\gamma}_k,\eta)$ in $\bm{\gamma}_k$, the KKT conditions are obtained as:
\begin{subequations}\label{eq:KKT}
	\begin{align}
	&\!\!\!\!\sum_{t}\mathbb{E}[{T_k^{\varphi_t}}^HT_k^{\varphi_t}]\bm{\gamma}_k \hspace{-0.5mm}+ \hspace{-0.5mm}\sum_{t}\mathbb{E}[{T_k^{\varphi_t}}^HZ_{k,NL}^{\varphi_t}] \hspace{-0.5mm}+\hspace{-0.5mm} \eta{T_k^{\varphi_0}}^H = 0, \!\!\!\!\\
	&\!\!T_k^{\varphi_0}\bm{\gamma}_k = T_k^{\varphi_0'},\!\!\!\!
	\end{align} 
\end{subequations}
where $T_k^{\varphi_0'}={\bm{\overline{h}}_{k}^{\varphi_0}}^T\bm{\overline{W}}_k \bm{\overline{\tilde{\phi}}}_{k0}^0\bm{\overline{\Psi^{'}}}\;\bm{\overline{\Phi^{''}}}_k$. Using~\eqref{eq:KKT}, the optimal solution $\bm{\widehat{\gamma}}_k$ and corresponding Lagrangian multiplier $\widehat{\eta}$ are:
\begin{subequations}
	\begin{align}\nonumber
	\bm{\widehat{\gamma}}_k =&\textstyle  \widehat{\eta}\left(\sum_{t}\mathbb{E}[{T_k^{\varphi_t}}^HT_k^{\varphi_t}]\right)^{-1}{T_k^{\varphi_0}}^H-\left(\sum_{t}\mathbb{E}[{T_k^{\varphi_t}}^HT_k^{\varphi_t}]\right)^{-1}\\
	&\textstyle \times \sum_{t}\mathbb{E}[{T_k^{\varphi_t}}^HZ_{k,NL}^{\varphi_t}],\\
	\widehat{\eta} = &\textstyle \frac{T_k^{\varphi_0'}+T_k^{\varphi_0}(\sum_{t}\mathbb{E}[{T_k^{\varphi_t}}^HT_k^{\varphi_t}])^{-1}\sum_{t}\mathbb{E}[{T_k^{\varphi_t}}^HZ_{k,NL}^{\varphi_t}] }{T_k^{\varphi_0}(\sum_{t}\mathbb{E}[{T_k^{\varphi_t}}^HT_k^{\varphi_t}])^{-1}{T_k^{\varphi_0}}^H}.
	\end{align}
\end{subequations}

\section{Numerical Experiment and Conclusion}\label{sec:num_exp} 
\label{page:R3C4} To evaluate the performance, we use the set of $16$ PA memoryless polynomial models obtained by measuring the outputs of 16 HMC943APM5E PA ICs at 28.5 GHz to the OFDM input signal of $200$ MHz bandwidth. Although, the signal is wideband, the memoryless model provides an accuracy around $-23$ dB in normalized mean square error~\cite{bri10}.  Besides, \label{page:R1C3a} two subarrays, each comprising $16$ PAs, are considered. The \label{page:R1C3} PAs are arranged in a uniform linear array and the distance between the two adjacent antennas is half the operating wavelength. The beamforming and steering weight for a given azimuth angle is determined using the procedure in~\cite{khan11}. \label{page:R1C3b} The crosstalk between two adjacent antennas is $-10$ dB, whereas, it decays as a square of distance between other two antennas~\cite{bri10}. Using the measured output of HMC943APM5E PA ICs at 28.5 GHz with $-10$ dB crosstalk, we identified the dual-input memoryless model for each PA using LS estimation as follows. For a given input signal $\bm{\mathring{x}}_k$, a test crosstalk signal $\bm{\mathring{c}}_{kl}$ to the PA, and the measured output $\bm{\mathring{y}}_{kl}$, the coefficients of the PA are identified using LS estimation as: $\bm{\widehat{\Phi}}_{kl} = \bm{\Psi}(\bm{\mathring{x}}_k,\bm{\mathring{c}}_{kl})^\dagger\hspace{0.2mm} \bm{\mathring{y}}_{kl}$.

\begin{figure}[!t]
	\centering
	\!\!\subfigure[]{\includegraphics[width=1.8in]{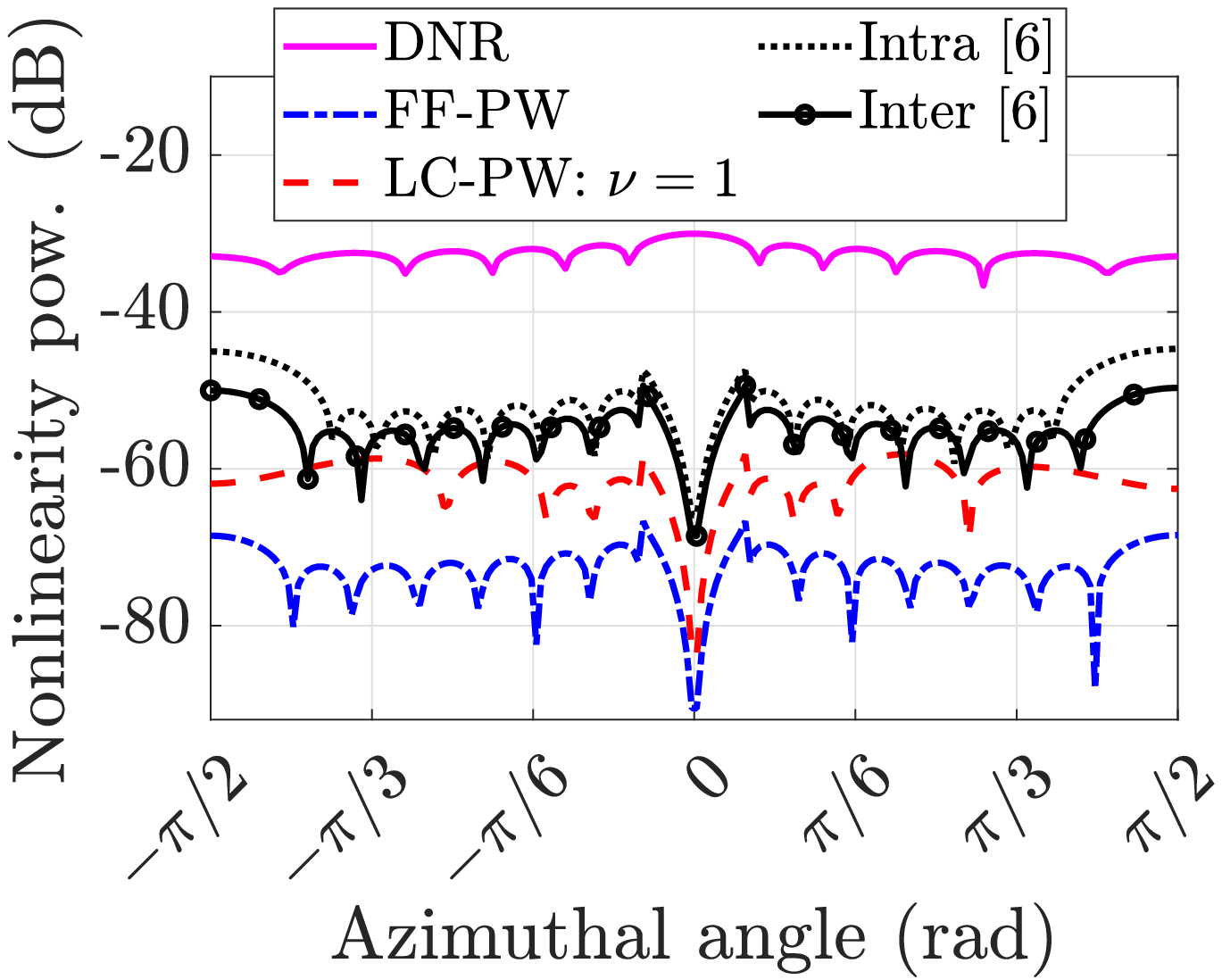}}\vspace{-1mm}\hspace{-3mm}
	\subfigure[]{\includegraphics[width=1.8in]{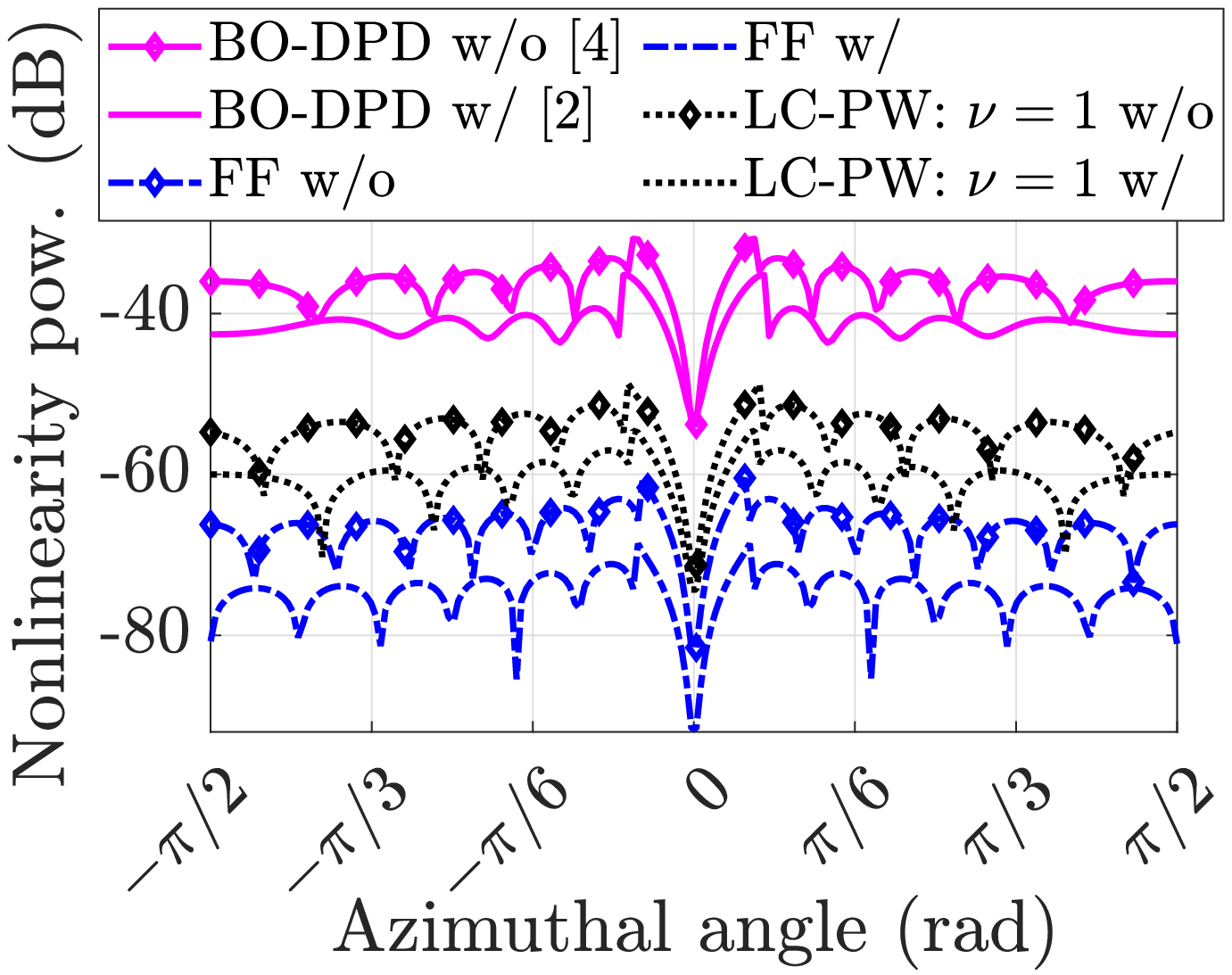}}\vspace{-1mm}\!\!\!\!
	\caption{\small Performance comparison of the different PW schemes against (a) the benchmark schemes and (b) the systems with only BO-DPD.}
	\label{fig:results}
\end{figure} 

\begin{table}[t!]
	\begin{center}
		\caption{Average ACPR of two adjacent channels.}
		\label{tab:table1}\footnotesize
		\begin{tabular}{|l|c|l|c|} 
			\hline
			Scheme & ACPR (dB) & Scheme & ACPR (dB)\\
			\hline
			Intra~\cite{yan07} & 41.6683 & BO-DPD w/ CTP~\cite{luo03} & 33.0106 \\
			\hline
			Inter~\cite{yan07} & 45.7127 & BO-DPD w/o CTP~\cite{ng06} & 27.5087\\
			\hline
			FF-PW & 63.4938 & LC-PW & 52.1726\\
			\hline
		\end{tabular}
	\end{center}
\end{table}
\label{page:R1C2} Fig.~\ref{fig:results} depicts the performance comparison of the schemes in the nonlinear radiations. A notch in each curve at $0$ rad is due to BO-DPD linearization in the direction. PW coefficients are optimized for the angle range $[-\pi/3, \pi/3]$ and $[-\pi/2, \pi/2]$ for Figs.~\ref{fig:results}(a) and \ref{fig:results}(b), respectively. So, intra-PW and inter-PW schemes overshoots beyond the range in  Fig.~\ref{fig:results}(a). In the direct nonlinear radiation (DNR), the message signal is directly transmitted from the subarray where the average power of the transmit signal is normalized to $0$ dB. So, the predistortion schemes give significant performance enhancement against DNR. Further, FF-PW and LC-PW provide the improvements by $20.01$ dB and $9.88$ dB against intra-PW due to their higher DOF in PW, but, inter-PW gives marginal improvement of $3.44$ dB. Fig~\ref{fig:results}(b) quantifies the performance improvement due to crosstalk preprocessing (CTP) in the schemes. Here, a scheme name followed by w/ (w/o) represents the predistortion with (without) CTP. In BO-DPD w/o~\cite{ng06} and BO-DPD w/~\cite{luo03}, only DPD is used to linearize the BO outputs. Thus, the PW schemes always perform better. Moreover, BO-DPD, LC-PW, and FF-PW schemes with CTP provide the respective on average improvements by $5.06$ dB, $5.82$ dB, and $8.11$ dB against the schemes without CTP. \label{page:R1C5} Also, the average adjacent channel power ratios (ACPRs) of the schemes are shown in Table~\ref{tab:table1}. Here, average ACPR is the average of the ACPRs of the two adjacent channels which are computed as in~\cite{bri10}. Again, the FF-PW has the best average ACPR while the LC-PW has an intermediate performance.

\makeatletter
\renewenvironment{thebibliography}[1]{%
	\@xp\section\@xp*\@xp{\refname}%
	\normalfont\footnotesize\labelsep .5em\relax
	\renewcommand\theenumiv{\arabic{enumiv}}\let\p@enumiv\@empty
	\vspace*{-1pt}
	\list{\@biblabel{\theenumiv}}{\settowidth\labelwidth{\@biblabel{#1}}%
		\leftmargin\labelwidth \advance\leftmargin\labelsep
		\usecounter{enumiv}}%
	\sloppy \clubpenalty\@M \widowpenalty\clubpenalty
	\sfcode`\.=\@m
}{%
	\def\@noitemerr{\@latex@warning{Empty `thebibliography' environment}}%
	\endlist
}
\makeatother

\bibliographystyle{IEEEtran}
\bibliography{references_PA_DPD}

\end{document}